\documentstyle[12pt,epsfig]{elsart}
\begin{document}
\begin{frontmatter}
\title{On the nuclear dependence of the $\mu^-\to e^-$ conversion branching ratio }

\author{T.S. Kosmas}
\address{Theoretical Physics Division,
University of Ioannina, GR-45110 Ioannina, Greece}

\begin{abstract}
The variation of the coherent branching ratio $R_{\mu e}$ (ratio
of the $\mu^-\to e^-$ reaction rate divided by the total
muon-capture rate) through the periodic table is studied by using
exact muon wave functions. It was found that, by using very heavy
nuclei (e.g. \nuc{197}Au, the SINDRUM II target) as $\mu^-\to e^-$
conversion stopping-targets, the above ratio is favored by a
factor of about four to five than by using light ones (e.g.
\nuc{48}Ti, chosen as PRIME target).
\end{abstract}

\begin{keyword}
rare muon decays \sep muon-to-electron conversion \sep muon
capture \sep lepton flavor violation

\PACS 23.40.Bw \sep 24.10.-i \sep 13.35.Bv \sep 12.60.Cn
\end{keyword}

\end{frontmatter}

\newcommand{\newc}{\newcommand}
\newc{\ra}{\rightarrow}
\newc{\lra}{\leftrightarrow}
\newc{\beq}{\begin{equation}}
\newc{\eeq}{\end{equation}}
\newc{\barr}{\begin{eqnarray}}
\newc{\earr}{\end{eqnarray}}

\def\mue{(\mu ^-,e^-)}
\def\al{\alpha}
\def\be{\beta}
\def\del{\delta}
\def\la{\lambda}
\def\ka{\kappa}

The present status of the exotic neutrinoless $\mu^-\to e^-$
conversion in nuclei,
\begin{equation}
\mu^-_b + (A,Z) \to e^- + (A,Z)^{\ast} \, ,
\label{mue}
\end{equation}
has comprehensively been discussed recently
\cite{Schaaf,Kuno,tsk01} both from experimental
\cite{Honec,Molzon,MECO,Wintz,NUFACT,PRIME} and theoretical
\cite{Wei-Fei,Ernst,Marci,KV88,Czarne,tskHaw} point of view.
Process (\ref{mue}) violates the $L_i$, ($i=\mu, e$) quantum
numbers and has been proposed \cite{Marci} as one of the best
probes to test the existence of charged-lepton flavor
conservation, among a great number of similar processes predicted
by modern gauge and supersymmetric theories.

Several experiments have been designed to explore process
(\ref{mue}) and performed at PSI (and earlier at TRIUMF) on
\nuc{48}Ti, \nuc{208}Pb and \nuc{197}Au targets
\cite{Schaaf,Honec,Wintz,NUFACT}. They have put so far, only
bounds on the branching ratio $R_{\mu e}$. Presently, the
published best upper limit is $R^{Ti}_{\mu e} \leq 6.1\times
10^{-13}$ \cite{Wintz}.
The ongoing SINDRUM II experiment (PSI) is now using \nuc{197}Au
as stopping target and the extracted preliminary limit constitutes
an improvement over the previous one set on a heavy target
(\nuc{208}Pb \cite{Honec}) by two orders of magnitude
\cite{Wintz,NUFACT}. The planned MECO experiment (Brookhaven) is
going to use \nuc{27}Al target and a very intense pulsing
muon-beam \cite{Molzon,MECO} to reach a sensitivity of roughly $
R^{Al}_{\mu e} \leq 2\times 10^{-17}$ \cite{Molzon,MECO} which
implies an improvement of the present limits by about three orders
of magnitude. It should be mentioned that very recently a new
$\mu^-\to e^-$ conversion experiment on \nuc{48}Ti (PRIME) was
announced to be performed at KEK \cite{Kuno,PRIME} aiming to push
the limit down to $R^{Ti}_{\mu e} \le 10^{-18}$. In all these
experiments, the signature of reaction (1) is a single electron
with energy $E_e = m_\mu-\epsilon_b-m_e$ (neglecting recoil) where
$\epsilon_b$ is the muon binding energy in the $1s$ orbit and
$m_\mu$ ($m_e$) the muon (electron) mass.

From the theoretical point of view
\cite{Marci,KV88,Czarne,tskHaw,Kos-Kov,Chiang}, process
(\ref{mue}) constitutes a very good interplay between atomic,
nuclear, particle and non-standard physics \cite{Czarne,tskHaw}.
The $\mu^-\to e^- $ conversion Hamiltonians which result in the
context of many extensions of the standard model proposed up to
now, in general, give rise to coherent and incoherent processes
\cite{Kos-Kov}. In several models, like those for which the
isoscalar couplings of the vector and scalar interactions are not
very small \cite{tsk01,Kos-Kov}, the coherent action of all
nucleons in $\mu^- \to e^-$ leads to enhancement of conversion
electrons thereby making it potentially a sensible indicator for
lepton flavor violation (LFV) effects
\cite{Wei-Fei,Ernst,Marci,KV88}. In addition, the $g.s.\to g.s.$
transitions are favored due to Pauli blocking effects which
prevent the formation of excited states. The present work is
motivated from the necessity to investigate the nuclear physics
aspects of process (1). We study the nuclear structure dependence
of the coherent branching ratio $R_{\mu e}$ throughout the
periodic table by performing exact calculations of the
muon-nucleus overlap integrals.


The expression for $R_{\mu e}$, to leading order in the
non-relativistic reduction, for the coherent process has been
written in the form \cite{Kos-Kov}
\begin{equation}
R_{\mu e} \, \equiv \, \frac{\Gamma_{\mu e}(A,Z)}{\Gamma_{\mu
c}(A,Z)} \, = \, \frac{G^2_F}{2 \pi } \, {\mathcal Q} \, \frac{ {\
p_e \ E_e} \, \vert {\mathcal M}_{V,S}^{(0)} \vert ^2}{\Gamma_{\mu
c}} \, , \label{Rme}
\end{equation}
where $\Gamma_{\mu e}$ stands for the $\mu^-\to e^-$ conversion
rate and $\Gamma_{\mu c}$ for the total rate of the ordinary
muon-capture, $\mu^-_b + (A,Z) \to \nu_\mu + (A,Z-1)$
\cite{Rosen,Suzu}. The factor $G_F^2/2$ corresponds to
non-photonic mechanisms and for photonic ones it should be
replaced by the ratio $(4\pi \alpha)^2/q^4 $, where $\alpha$ is
the fine structure constant. In Eq. (\ref{Rme}), $p_{e}$ denotes
the outgoing-electron momentum connected to the excitation energy
$E_x$ of the daughter nucleus ($E_x=E_f-E_{gs}$) through the
relation
\begin{equation}
p_e \approx q \, =\, m_\mu -\epsilon_b - E_x . \label{Kinem}
\end{equation}
$q =\vert\bf q\vert$ is the magnitude of the momentum transfer (we
neglect the electron mass). The quantity ${\mathcal Q}$ of Eq.
(\ref{Rme}) depends very weakly on the nuclear structure and, in
principle, it contains scalar (S), vector (V), axial-vector (A),
pseudo-scalar (P), and tensor (T) coupling terms \cite{Kos-Kov}.
Especially for photonic diagrams, which are the main concern of
this work, $\mathcal Q$ is rather nuclear-structure independent.
Hence, the main nuclear physics aspects of $R_{\mu e}$ are
accumulated in the last fraction of Eq. (\ref{Rme}).

The matrix elements ${\mathcal M}^{(\tau)}_\alpha$,
$\alpha=V,S,A,P,T$, which enter the expression of $R_{\mu e}$ are
defined by
\begin{equation}
{\mathcal M}^{(\tau)}_\alpha \, = \, \langle f|\sum_{j=1}^A
\Theta^{\tau}_{\alpha}( j) e^{-i{\bf q} \cdot {\bf r}_j} \, \Phi
({\bf r}_j) | i \rangle \, , \label{mue-ME}
\end{equation}
$\tau=0$ for isoscalar and $\tau=1$ for isovector operators,
respectively, where $\vert i \rangle$ the initial and $\vert f
\rangle$ the final nuclear state. $\Phi_{\mu} ({\bf r}_j)$
represents the muon wave function evaluated at the position of the
$j^{th}$ target-nucleon. The functions $\Theta^{\tau}_{\alpha}(j)$
contain the spin-isospin dependence of the $\mu^-\to e^-$ operator
\cite{tsk01}. For the coherent process ($\vert f\rangle = \vert
i\rangle$), in the case of scalar and vector interactions,
${\mathcal M}^{(\tau)}_{V,S}$ are written in terms of the
ground-state proton, neutron densities as
\begin{equation}
{\mathcal M}^{(\tau)}_{V,S}(q)=\int \left[ \rho_{p}({\bf r}) \pm
\rho_{n} ({\bf r}) \right] e^{- i {\bf q} \cdot {\bf r}}
\Phi_{\mu} ({\bf r}) d^3 {\bf r} \equiv {\mathcal F}_p(q) \pm
{\mathcal F}_n(q) \label{M-scal} \, ,
\end{equation}
the (+) sign corresponds to $\tau=0$ and the (-) to $\tau=1$
channel, where
\begin{equation}
{\mathcal F}_{p,n} (q) = \int \rho_{p,n} ({\bf r}) \; e^{- i {\bf
q} \cdot {\bf r}}\; \Phi_{\mu} ({\bf r}) d^3 {\bf r} \label{Fpn} \, .
\end{equation}
The proton (neutron) density $\rho_p$ ($\rho_n$) is normalized to
the atomic number Z (neutron number N) of the nucleus in question.
For our purposes here, the required densities $\rho_{p}$ are taken
from experiment \cite{Vries}. For photonic mechanisms only protons
of the target-nucleus contribute and hence, ${\mathcal
M}^{(0)}_{V,S}={\mathcal M}^{(1)}_{V,S}={\mathcal F}_{p}(q)$.

For light and medium nuclei (see discussion of the results below)
${\mathcal M}^{(\tau)}_\alpha$ can be reliably evaluated in a
straightforward way by factorizing outside the integrals of Eq.
(\ref{mue-ME}) a suitably averaged muon wave function
$\langle\Phi^{1s}_\mu\rangle$. Under these conditions Eq.
(\ref{mue-ME}) is approximated by
\begin{equation}
{\overline {\mathcal M}}^{(\tau)}_\alpha \, = \,
\langle\Phi^{1s}_\mu \rangle \langle f\mid \sum_{j=1}^A
\Theta^{\tau}_{\alpha}(j) e^{-i{\bf q} \cdot {\bf r}_j} \mid
i\rangle \, \equiv \, \langle\Phi^{1s}_\mu \rangle \,
M^{(\tau)}_\alpha \, , \label{mean-Phi}
\end{equation}
where $M^{(\tau)}_\alpha$ involve the pure nuclear physics aspects
of the $\mu^-\to e^-$ process. Equation (\ref{mean-Phi}) for
photonic diagrams,  is written as ${\overline {\mathcal
M}}_{V,S}^{(0,1)} \equiv {\overline {\mathcal
F}}_{p}(q)=\langle\Phi^{1s}_\mu\rangle \, Z F_{p}(q)$. For the
mean muon wave function $\langle\Phi^{1s}_{\mu}\rangle$, a
simplified expression (see e.g. Ref. \cite{Rosen} and Eqs. (22),
(23) of Ref. \cite{Chiang}) was used in muon capture studies by
many authors \cite{Rosen}. In previous estimations of the
branching ratio $R_{\mu e}$, the same expression for
$\langle\Phi^{1s}_\mu\rangle$ was adopted \cite{KV88} for both the
numerator and the denominator of Eq. (\ref{Rme}) in order to
reduce the uncertainties inserted via the use of Eq.
(\ref{mean-Phi}) in $\mu^-\to e^-$ and $\mu^-\to \nu_\mu$
processes.

In the special case of $g.s.\to g.s.$ transitions for spin zero
(J=0) light nuclei, $M^{(\tau)}_{V,S}$ are determined by the
elastic scattering (monopole) nuclear form factors $F_p$, $F_n$
\cite{KV88} and they are simply given by
$M^{(\tau)}_{V,S} (q) \, = \, Z F_{p} (q) \pm N {F}_{n} (q)$.
In the general case of nuclei with ground-state spin $J \ne 0$
(e.g. \nuc{27}Al, the MECO target, with $J=\frac{5}{2}$),
$M^{(\tau)}_{V,S}$ contain, in addition to monopole ($L=0$) form
factors, contributions arising from other multipoles ($L=2, 4,
...)$. However, the latter contributions are not significant
\cite{tsk01}.


The main goal of the present work, was to study systematically the
exact nuclear-structure dependence of $R_{\mu e}(A,Z)$ by
computing the integrals of Eq. (\ref{M-scal}) for the dominant
coherent $\mu^- \to e^-$ conversion and investigate the influence
on $R_{\mu e}$ of the following effects: (i) the approximate
evaluation of the muon-nucleus overlap integrals [see Eq.
(\ref{mean-Phi})] and (ii) the neglect of the muon-binding energy
$\epsilon_b$ in Eq. (\ref{Kinem}). The latter assumption implies
that for all nuclei in the coherent mode we have
\begin{equation}
q=p_{e} \approx m_{\mu}/c \ = \ 0.534 \ fm^{-1} \, ,
\qquad E_e \ \approx \ 105.6 \ MeV \, .
\label{Noeb}
\end{equation}
In this work, the rather simple photonic mechanism, where only the
target-protons contribute to $R_{\mu e}$, was examined. The main
steps followed in the calculational procedure and the results
obtained are briefly discussed below.

In the first step, the variation of the quantity $Z\vert
F_Z(q)\vert^2$ through the periodic table was studied. As is well
known, Weinberg and Feinberg \cite{Wei-Fei} using the above
approximations have shown that, for the coherent $\mu^-\to e^-$
rate it holds
\begin{equation}
R_{\mu e} \, \propto \, Z\vert F_{NN} (q)\vert^2. \label{Gme-prop}
\end{equation}
In Fig. 1(a) the results obtained for Eq. (\ref{Gme-prop}) in the
following cases are illustrated:

(i) In the first case, experimental form factors
$F_Z(q)=F_{NN}(q)$ \cite{Vries} were used at the values of $q$
given by Eq. (\ref{Noeb}) (dashed-dotted line) and by Eq.
(\ref{Kinem}) (dotted line).

(ii) In the second case, the phenomenological expression for
$F_{NN}(q)$ [see Eq. (16) of Ref. \cite{Wei-Fei}] was used in Eq.
(\ref{Gme-prop}) (solid line).

The common feature of the three curves in Fig. 1(a) is the fact
that they show a maximum in the region of $A\approx 60$ (copper
region) as had been estimated in Ref. \cite{Wei-Fei}. This
behavior was generally adopted by the experimentalists exploring
the $\mu^-\to e^-$ process \cite{Schaaf,Kuno}. As it can be seen,
in the region of light nuclei, the three curves nearly coincide,
which means that the approximation of Eq. (\ref{Noeb}) is
reasonable in this region, but for medium and heavy nuclei where
$\epsilon_b$ becomes significant, the obtained rates when $q$ is
given by Eq. (\ref{Kinem}) are larger by about a factor of two
than those when $q$ is given by Eq. (\ref{Noeb}).

In the second step of the calculations, the variation of the
quantity $R_{\mu e}/{\mathcal Q}_{ph}$ [see Eq. (\ref{Rme})],
which contains the main nuclear-structure dependence of $R_{\mu
e}$, was studied. Two cases were distinguished:

(i) In the first case, the mean muon wave function was inserted in
Eq. (\ref{Rme}). This gives
\begin{equation}
R_{\mu e} \, \propto \, {16\pi\alpha^2} \,
\frac{ {\ p_e \ E_e} }{q^4} \, \frac{ \langle \Phi_\mu \rangle ^2 \,
Z^2 \, \vert F_{p}(q) \vert ^2}{\Gamma_{\mu c}} \,
\label{Rme-Qph-eff}
\end{equation}
The results obtained in this way, are presented in Fig. 1(b) and
correspond to the two choices of coherent momentum transfer
discussed before: (a) by using the values of Eq. (\ref{Noeb})
(solid line) and (b) by using the values of $q$ given by Eq.
(\ref{Kinem}) ($E_x=0$ for coherent mode) (dashed line). In choice
(a) we see that $R_{\mu e}(A,Z)$ presents a maximum at the region
of $A\approx 130$ which is not in accordance with the estimation
of Ref. \cite{Wei-Fei}. This is due to the fact that in Ref.
\cite{Wei-Fei} the gross nuclear dependence of $\Gamma_{\mu c}$
was considered as linear in Z which is equivalent to a constant
Primakoff function $f_{GP}(A,Z)$. It is well known, however, that
$f_{GP}(A,Z)$ is strongly dependent on the mass excess (see e.g.
Ref. \cite{KV88}). In this work, in order to minimize such
uncertainties, experimental data for the total muon capture rates
$\Gamma_{\mu c}$ were used \cite{Suzu}. From the results of choice
(b) we see that $R_{\mu e}(A,Z)$ shows a slow increase up to the
heaviest nuclei.

(ii) In the second case, the explicit muon-nucleus overlap
integral of Eq. (\ref{mue-ME}) was used (numerical integration) in
Eq. (\ref{Rme}). Then, the variation of $R_{\mu e}(A,Z)$ relies on
the expression
\begin{equation}
R_{\mu e} \, \propto \, {16\pi\alpha^2} \,
\frac{p_e \ E_e}{q^4} \, \frac{\vert {\mathcal F}_{p} \vert
^2}{\Gamma_{\mu c}} \, \label{Rme-Qph-ex}
\end{equation}
This requires the use of the exact muon wave function calculated
as in Ref. \cite{Kos-Lag}. Here the values of $q$ are provided by
Eq. (\ref{Kinem}). The results are represented by stars ($*$) in
Fig. 1(b). We see that, the exact evaluation of the muon-nucleus
overlap integrals of Eq. (\ref{M-scal}), shows linear increase of
$R_{\mu e}$ as function of A (or Z \cite{Kos-Lag}). For heavy and
very heavy nuclei (region of $^{197}$Au and $^{208}$Pb), the use
of the exact muon wave function in Eq. (\ref{Rme-Qph-ex}), gives
much larger rates than the approximation of the averaged muon wave
function Eq. (\ref{Gme-prop}). The comparison is much worse if
$\epsilon_b$ is neglected in the kinematics. One must notice that,
the presence of the factor $q^{-4}$ in Eq. (\ref{Rme-Qph-ex})
(photonic mechanism), inserts an additional (A,Z) dependence on
the branching ratio $R_{\mu e}$ which had been previously
overlooked.


In conclusion, the gross variation of $R_{\mu e}$ obtained by the 
exact results, shows a linear rise with A which can be attributed 
to the coherent effect. Due to this behaviour of $R_{\mu e}(A,Z)$, 
very heavy nuclei are favored by a factor of about five to be used 
as $\mu^- \to e^-$ conversion targets. It is worth mentioning that, 
the above factor may, in some cases, be partly compensated by other 
experimental advantages of some specific muon-stopping targets 
\cite{Molzon}. We also remark that in the present work we neglected 
relativistic atomic effects related to $\mu^-\to e^-$ process 
\cite{Czarne} which might influence a bit (depending on the nucleus) 
the results for $R_{\mu e}$.


In summary, we have investigated the dependence of the $\mu^-\to
e^-$ conversion branching ratios on the nuclear parameters A and Z
throughout the periodic table. This exotic process is an
interesting and important one to be studied, since stringent
bounds for the charged-lepton flavor violating parameters already
exist and significant improvements over these limits are feasible
in the not-too-distant future from the SINDRUM II, MECO and PRIME
experiments. We performed direct calculations of the muon-nucleus
overlap integrals (for photonic mechanisms) and found that $R_{\mu
e}(A,Z)$ keeps increasing up to the very heavy nuclei. This shows
that, by using such isotopes (e.g. \nuc{197}Au, present SINDRUM II
target) as targets in $\mu^- \to e^-$ conversion experiments,
$R_{\mu e}$ is favored by a factor of about four to five than by
using light isotopes. Finally, the present exact results were
exploited to test the validity of previous approximations made on
the study of the $\mu -e$ conversion and shed more light in this
situation.

\bigskip
The author wishes to thank Dr. Y. Kuno for financial support and
Dr. Andries van der Schaaf for fruitful discussions on SINDRUM II
experiments.


\vspace*{2.0cm}

\centerline{\bf FIGURE CAPTION}

\vspace*{0.4cm}

\noindent {\bf Fig. 1. }{ Variation of the $\mu^-\to e^-$
conversion branching ratio, $R_{\mu e}$, through the periodic
table assuming that the nuclear-structure dependence of $R_{\mu
e}$ is described: (a) by Eq. (\ref{Gme-prop}) \cite{Wei-Fei} and
(b) by Eqs. (\ref{Rme-Qph-eff}) and (\ref{Rme-Qph-ex}). For
details see the text.}


\begin{figure}[t]
\centerline{\epsfig{figure=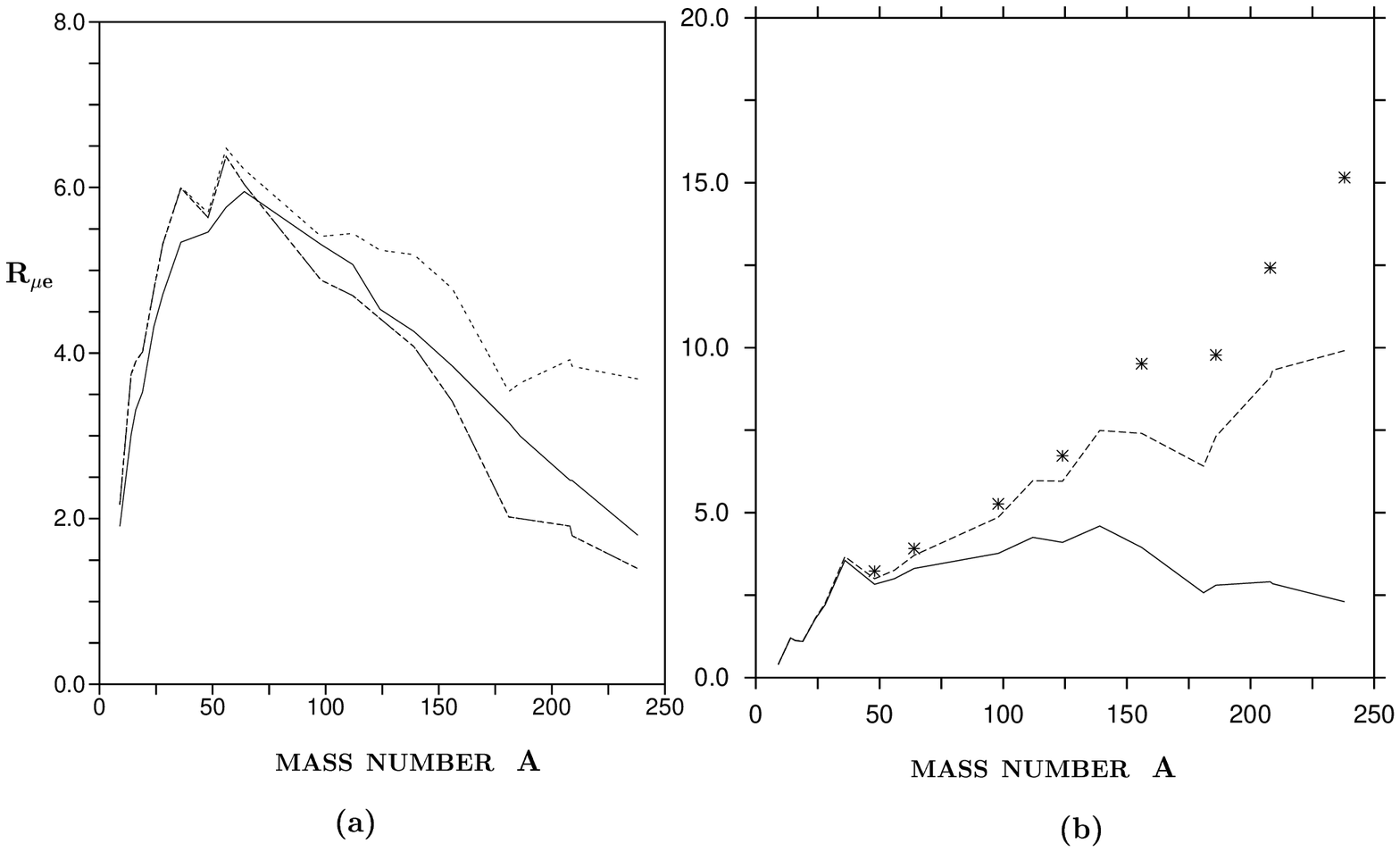}}
\caption{Variation of the$\mu^-\to e^-$ conversion branching
ratio, $R_{\mu e}$, through the periodic table assuming that the
nuclear-structure dependence of $R_{\mu e}$ is described: (a) by
Eq. (\ref{Gme-prop}) \cite{Wei-Fei} and (b) by Eqs.
(\ref{Rme-Qph-eff}) and (\ref{Rme-Qph-ex}). For details see the
text.}
\end{figure}

\end{document}